\documentclass[twocolumn,preprintnumbers,amsmath,amssymb]{revtex4}
\usepackage{graphicx}
\usepackage{dcolumn}
\usepackage{bm}

\begin{document}

\title{A Harris-Todaro Agent-Based Model to Rural-Urban Migration}

\author{Aquino L. Esp\'{\i}ndola}
 \affiliation{Instituto de F\'{\i}sica, Universidade Federal Fluminense\\
24.210-340, Niter\'oi, RJ, Brazil\\
aquino@if.uff.br}

\author{Jaylson J. Silveira}
\affiliation{Depto de Economia, Universidade Estadual Paulista - UNESP\\
14.800-901, Araraquara, SP, Brazil\\
jaylson@fclar.unesp.br}

\author{T. J. P. Penna}
\affiliation{Instituto de F\'{\i}sica, Universidade Federal Fluminense\\
24.210-340, Niter\'oi, RJ, Brazil\\
tjpp@if.uff.br\\}

\date{\today}

\begin{abstract}
The Harris-Todaro model of the rural-urban migration process is revisited under an agent-based approach. The migration of the workers is interpreted as a process of social learning by imitation, formalized by a computational model. By simulating this model, we observe a transitional dynamics with continuous growth of the urban fraction of overall population toward an equilibrium. Such an equilibrium is characterized by stabilization of rural-urban expected wages differential (generalized Harris-Todaro equilibrium condition), urban concentration and urban unemployment. These classic results obtained originally by Harris and Todaro are emergent properties of our model.
\end{abstract}

\maketitle

\section{Introduction}
\label{introduction}

In this paper we turn upon the seminal Harris and Todaro \cite{harristodaro} work, which  together with Todaro \cite{todaro} is considered one of the starting points of the classic rural-urban migration theory \cite{ray}. The hypothesis and predictions of Harris-Todaro model have been subjected to econometric evaluation and have been corroborated by several studies \cite{yap, mazumdar, willianson, ghatak}. The key hypothesis of Harris and Todaro are that migrants react mainly to economic incentives, earnings differentials, and the probability of getting a job at the destination have influence on the migraton decision. In other words, these authors posit that rural-urban migration will occur while the urban expected wage \footnote{The urban sector wage times the probability of getting a job in this sector.} exceed the rural wage. From this {\it crucial assumption}, as denominated by Harris-Todaro \cite{harristodaro}, is deduced that the migratory dynamics leads the economic system toward an equilibrium with urban concentration and high urban unemployment.

In our previous works \cite{nos,nos2} we analyzed the rural-urban migration by means of an agent-based computational model taking into account the influence of the neighborhood in the migration decision. The inclusion of the influence of neighbors was done via an Ising like model. The economic analogous  to the external field in the Ising hamiltonian was the differential of expected wages between urban and rural sectors. Therefore, in theses works \cite{nos, nos2} the crucial assumption of Harris and Todaro were taken for granted.

Now, we are motivated by the following question: can the crucial assumption and equilibrium with urban concentration and urban unemployment obtained from the original Harris-Todaro model be generated as emergent properties from the interaction among adaptative agents? In order to answer this question we implemented an agent-based computational model in which workers grope for best sectorial location over time in terms of earnings. The economic system simulated is characterized by the assumption originally made by Harris and Todaro.

The paper is arranged as follows. Section \ref{haristodaromodel} describes the analytical Harris-Todaro model showing its basic equilibrium properties. In Section \ref{harristodaroagentbasedmodel} we present the implementation of the computational model via an agent-based simulation and compare its aggregate regularities with the analytical equilibrium properties. Section \ref{conclusion} shows concluding remarks.

\section{The Harris-Todaro Model}
\label{haristodaromodel}

\subsection{Assumptions}
\label{assumptions}

Harris and Todaro \cite{harristodaro} studied the migration of workers in a two-sector economic system, namely, rural sector and urban sector. The difference between these sectors are the type of goods produced, the technology of production and the process of wage determination. The rural sector is specialized in the production of agricultural goods. The productive process of this sector can be described by a Cobb-Douglas production function:

\begin{equation}
Y_a=A_aN_a^\phi,
\label{ya}
\end{equation}
where $Y_a$ is the production level of the agricultural good, $N_a$ is the amount of workers used in the agricultural production, $A_a>0$ and $0<\phi <1$ are parametric constants.

Similarly, the urban sector also has its productive process described as Cobb-Douglas production function:\footnote{Harris and Todaro set their model using sectorial production function with general functional form, i.e., in each sector $i$, $Y_i=f_i(N_i)$ with $f_i^\prime (N_i)>0$ and $f_i^{\prime\prime} (N_i)<0$. Except where it is indicated, the results presented in this section are valid for this general case. The Cobb-Douglas form is a standard assumption about technology.}

\begin{equation}
Y_m=A_mN_m^\alpha,
\label{ym}
\end{equation}
where $Y_m$ is the production level of the manufactured good, $N_m$ is the quantity of workers employed in the production of manufactured goods, $A_m>0$ and $0<\alpha <1$ are parametric constants.\footnote{The endowment of land of the rural sector and the stocks of capital of both sectors are given for the period of analysis.}

Both goods and labor markets are perfectly competitive. Nevertheless, there is segmentation in the labor market due to a high minimum urban wage politically determined.

In the rural sector, the real wage, perfectly flexible, is equal to the marginal productivity of labor in this sector \footnote{This marginal product is the derivative of the agricultural production function, eq. (\ref{ya}), with respect to $N_a$ multiplied by $p$.}:

\begin{equation}
w_a=\phi A_aN_a^{\phi - 1}p,
\label{wa}
\end{equation}
where $w_a$ is the real wage and $p$ is the price of the agricultural good, both expressed in units of manufactured good. 
 
In the urban sector, a minimum wage, $w_m$, is assumed fixed institutionally at a level above equilibrium in this labor market. It can be formalized as \footnote{The right-hand side of this equation is the marginal productivity of the manufacturing labor, i.e., the derivative of manufacturing production function, eq. (\ref{ym}), with respect to $N_m$.}

\begin{equation}
w_m=\alpha A_mN_m^{\alpha-1},~~\mbox{such that}~~N_m\leq N_u,
\label{wm}
\end{equation}
where $N_u$ is the amount of workers in the urban sector.

The relative price of the agricultural good in terms of the manufactured good, $p$, varies according to the relative scarcity between agricultural and manufacturated goods. Then,\footnote{Actually, Harris and Todaro worked with a general form, $\rho$ denotes a function in their work not a constant value as used by us.}

\begin{equation}
p=\rho\left(\frac{Y_m}{Y_a}\right)^\gamma,
\label{p}
\end{equation}
where $\rho>0$ and $\gamma>0$ are a parametric constants. $\gamma$ is the elasticity of $p$ with respect to the ratio $Y_m/Y_a$.

The overall population of workers in the economy is $N$, which is kept constant during the whole period of analysis. By assumption there are only two sectors and rural prices are wholly flexible, which implies that there is full employment in the rural area, i.e., all workers living at the rural sector are employed at any period. Then at any period the following equality is verified:

\begin{equation}
N_a+N_u=N.
\label{nanu}
\end{equation}

\subsection{Temporary Equilibrium}
\label{temporaryequilibrium}

Given a parametric constant vector $(A_a,A_m,\phi,\alpha,\rho,\gamma)$, an initial urban population $N_u$, and a minimum wage $w_m$ one can calculate the temporary equilibrium of the economic system by using eqs. (\ref{ya}-\ref{nanu}). 

From eq. (\ref{wm}) one can find the employment level at the manufacturing sector

\begin{equation}
N_m=\left(\frac{\alpha A_m}{w_m}\right)^{\frac{1}{1-\alpha}}.
\label{nm}
\end{equation}

Replacing eq. (\ref{nm}) in eq. (\ref{ym}) we get the production level of the manufacturing sector

\begin{equation}
Y_m=A_m^{\frac{1}{1-\alpha}}\left(\frac{\alpha}{w_m}\right)^{\frac{\alpha}{1-\alpha}}.
\label{ymfunc}
\end{equation}

From eq. (\ref{nanu}) one can obtain the relation

\begin{equation}
 N_a=N-N_u,
\label{na}
\end{equation}
which is used  with eq. (\ref{ya}) to obtain the agricultural production

\begin{equation}
Y_a=A_a\left(N-N_u\right)^\phi.
\label{yafunc}
\end{equation}

By using eqs. (\ref{p}), (\ref{ymfunc}) and (\ref{yafunc}) the terms of trade are determined

\begin{equation}
p=\rho\left[\frac{A_m^{\frac{1}{1-\alpha}}\left(\frac{\alpha}{w_m}\right)^{\frac{\alpha}{1-\alpha}}}{A_a(N-N_u)^\phi}\right]^\gamma.
\label{pfunc}
\end{equation}

Finally, by using eqs. (\ref{wa}), (\ref{na}) and (\ref{pfunc}), the rural wage in units of manufacturated good is obtained

\begin{equation}
w_a=\phi\rho A_a^{1-\gamma} A_m^{\frac{\gamma}{1-\alpha}}\left(\frac{\alpha}{w_m}\right)^{\frac{\alpha\gamma}{1-\alpha}}\frac{1}{(N-N_u)^{1-\phi+\phi\gamma}}.
\label{wafunc}
\end{equation}

In sum, the vector $(N_m,Y_m,N_a,Y_a,p,w_a)$ configures a temporary equilibrium that might be altered whether occurs a migration of workers, induced by the differential of sectorial wages, which changes the sectorial distribution of overall population.

\subsection{The Long Run Equilibrium}
\label{thelongrunequilibrium}

Harris and Todaro, in determining the long run equilibrium, i.e., the absence of a net rural-urban migratory flow, argue that the rural workers, in their decision on migrating to the urban area, estimate the expected urban wage, $w_u^e$, defined as:

\begin{equation}
w_u^e=\frac{N_m}{N_u}w_m.
\label{wu}
\end{equation}
The ratio $N_m/N_u$, which is the employment rate, is an estimative of the probability that a worker living at urban sector gets a job in this sector.

  As mentioned before, the key assumption of the model of Harris and Todaro is that there will be a migratory flow from the rural to the urban sector while the expected urban wage is higher than the rural wage. Thus, the long run equilibrium is attained when the urban worker population reaches a level such that the expected urban wage equates the rural wage:

\begin{equation}
w_u^e-w_a=0.
\label{wuwa}
\end{equation}

This equality is known in the economic literature as the {\it Harris-Todaro condition}.  Harris and Todaro argue that the differential of expected wages in eq. (\ref{wuwa}) can be a constant value $\delta\neq 0$. When this differential reaches $\delta$, the net migration ceases. This {\it generalized Harris-Todaro condition} can be expressed as follows:

\begin{equation}
w_u^e-w_a=\delta.
\label{wuwadelta}
\end{equation}

The level of the urban population that satisfies the eq. (\ref{wuwadelta}), i.e., the equilibrium urban share $n_u^*=N_u^*/N$, is determined from the solution of the equation resulting from substitution of equations (\ref{wafunc}), (\ref{wu}) in eq. (\ref{wuwadelta}):

$$
\frac{N_m}{N_u}w_m-\phi\rho A_a^{1-\gamma} A_m^{\frac{\gamma}{1-\alpha}}\left(\frac{\alpha}{w_m}\right)^{\frac{\alpha\gamma}{1-\alpha}}\times
$$

\begin{equation}
\frac{1}{(N-N_u)^{1-\phi+\phi\gamma}}=\delta.
\label{geneq}
\end{equation}
The solution of eq. (\ref{geneq}) is parametrized by the vector $(A_a,A_m,\rho,\gamma,\alpha,\phi,w_m)$.

Harris and Todaro \cite{harristodaro}, in order to evaluate the stability of the long run equilibrium, postulate a mechanism of adjustment that is based on the following function of sign preservation:

\begin{equation}
\dot N_u=\psi (w_u^e-w_a),~~\mbox{with}~~\psi^\prime(\centerdot)>0~~\mbox{and}~~\psi(0)=0.
\label{dndt}
\end{equation}

The differential equation that governs the state transition in the model of Harris and Todaro is obtained by replacing equations (\ref{wafunc}), (\ref{wu}) in eq. (\ref{dndt}). Based on this postulated adjustment process, Harris and Todaro \cite{harristodaro} show that the long run equilibrium is globally asymptotically stable. This means that the economy would tend to long run equilibrium with unemployment in the urban sector generated by the presence of a relatively high minimum wage for all possible initial conditions. From now on we will refer to the long run equilibrium simply as equilibrium.

\begin{figure}[hbt]
\includegraphics[width=7.5cm]{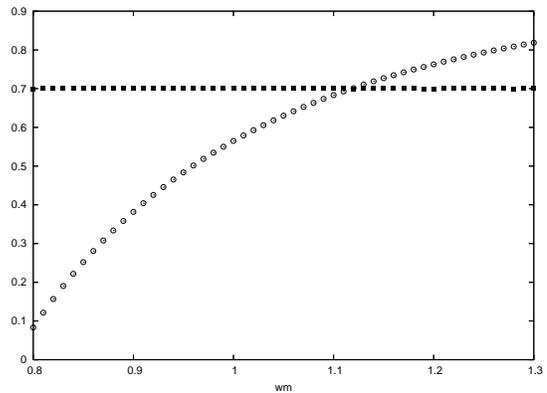}
\caption{Numerical solution of eq. (\ref{geneq}) for different values of $w_m$. Squares: urban share $n_u^*$; Circles: urban unemployment rate $(1-N_m/N)$. Fixed parameters used are $A_a=1.0$, $A_m=1.0$, $\phi=0.3$, $\alpha=0.7$, $\rho=1.0$ and $\gamma=1.0$.}
\label{nuwmn}
\end{figure}

Based on the numerical solutions of eq. (\ref{geneq}) one can evaluate the impact that the variation of the minimum wage and the elasticity of the terms of trade on the equilibrium. In Figure \ref{nuwmn} we see that under the hypothesis of a Cobb-Douglas technology, the equilibrium urban share, $n_u^*$, does not depend on the minimum wage $w_m$. However, changes in the value of $w_m$ reduces the labor demand on the manufacturing sector what results in higher unemployment rates in the equilibrium.

\begin{figure}
\includegraphics[width=7.5cm]{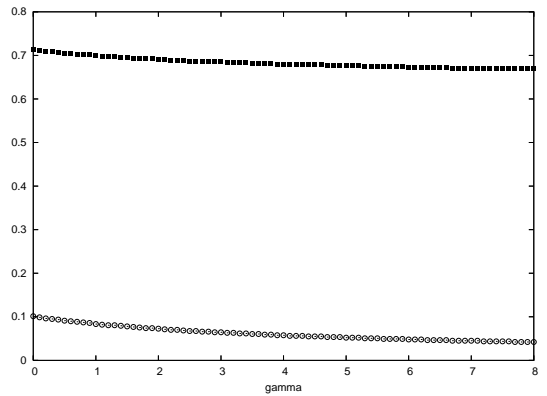}
\caption{Numerical solution of eq. (\ref{geneq}) for different values of $\gamma$. Squares: urban share $n_u^*$; Circles: urban unemployment rate, $(1-N_u^*/N)$. Fixed parameters used are $A_a=1.0$, $A_m=1.0$, $\phi=0.3$, $\alpha=0.7$, $\rho=1.0$ and $w_m=0.8$ }
\label{nugamman}
\end{figure}

In turn, as seen in Figure \ref{nugamman}, changes in the elasticity of the terms of trade alter slightly the equilibrium urban share and unemployment rate. A net migration toward urban sector shift the terms of trade to higher values. The greater $\gamma$ the greater this shift, what cause an increase in the rural wage in units of manufacturing good, becoming the urban sector less attractive.

\section{Harris-Todaro Agent-Based Model}
\label{harristodaroagentbasedmodel}

In this section we describe the implementation of the computational model we proposed, as well as the aggregate patterns obtained numerically and the comparison with the respective analytical results.

\subsection{Computational Implementation}
\label{computationalimplementation}

Initially, workers are randomly placed in a square lattice with linear dimension $L=500$. The reference values of the parameters used for these simulations are the same done to evaluate the equilibrium of the Harris-Todaro model, namely, $A_a=1.0$, $A_m=1.0$, $\phi=0.3$, $\alpha=0.7$, $\rho=1.0$ and $\gamma=1.0$. The value of the minimum wage used is $w_m=0.8$ and the initial urban fraction of the total population is $n_u=0.2$, where $n_u=N_u/N$ is the normalized urban population also called urban share. The initial value $n_u=0.2$ is in agreement with historical data of developing economies. Given these parameters, one can calculate the vector which characterizes temporary equilibrium of the system by using eqs. (\ref{nm}-\ref{wafunc}).

By using eq. (\ref{nm}), the employment level of the urban sector, $N_m$, is obtained. If $n_u\leq N_m/N$ all workers in the urban sector are employed and each individual $i$ earns the wage given by the manufacturing marginal productivity, $w_i=\alpha A_m N_u^{\alpha-1}$. Otherwise, $n_u>N_m/N$ there will be a fraction of $N_m/N_u$ workers employed, which earn the minimum wage, $w_i=w_m$, and $(1-N_m/N_u)$ workers unemployed, which earn a wage $w_i=0$.

Each worker can be selected to review his sectorial location with probability $a$, called activity \cite{thadeu}. Therefore, in each time step only a fraction of workers becomes potential migrants, going through the sectorial location reviewing process. Potential migrants will determine their satisfaction level of being in the current sector by comparing their earnings, $w_i$, among nearest neighbors. 

The potential migrant starts the comparison process with a initial satisfaction level $s_i=0$. When  $w_i>w_{neighbor}$ the satisfaction level $s_i$ is added in one unit; if $w_i<w_{neighbor}$, $s_i$ is diminished in one unit; if $w_i=w_{neighbor}$, $s_i$ does not change. After the worker has passed through the reviewing process his/her satisfaction level is checked. The migration will occur only if $s_i<0$, what means that the worker's $i$ earnings is less than the most of his/her nearest neighbors.

After all the potential migrants complete the reviewing process and have decided migrate or not, a new configuration of the system is set. Therefore, once again a new temporary equilibrium of the system is calculated by using eqs. (\ref{ymfunc}-\ref{wafunc}). The whole procedure is repeated until a pre-set number of steps is reached. It is important to emphasize that $N_m$ is kept constant throughout the simulation. Its given by eq. (\ref{nm}) which depends on the technological parameters, $\alpha,Am$, and the minimum wage, $w_m$, which are constants too.

\subsection{Analysis of the Emergent Properties}
\label{analysisoftheemergentproperties}

In this section we develop the analysis of the long run aggregate regularities of Harris-Todaro agent-based computational model. These long run properties will be compared between the solution of the analytical model and simulations we ran.

\begin{figure}[hbt]
\includegraphics[width=7.5cm]{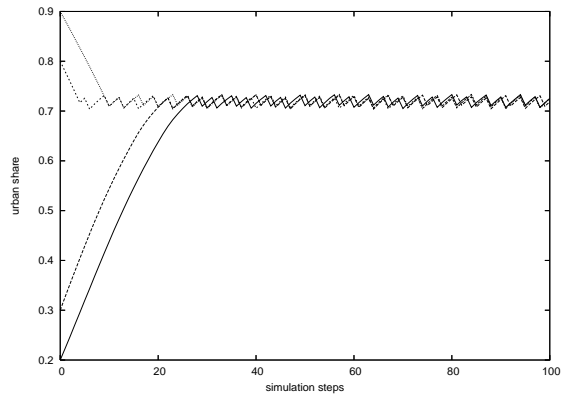}
\caption{Urban share $n_u$ as function of simulation steps. From top to bottom the initial urban shares are $0.9,~0.8,~0.3,~0.2$.}
\label{nut}
\end{figure}

\begin{figure}[hbt]
\includegraphics[width=7.5cm]{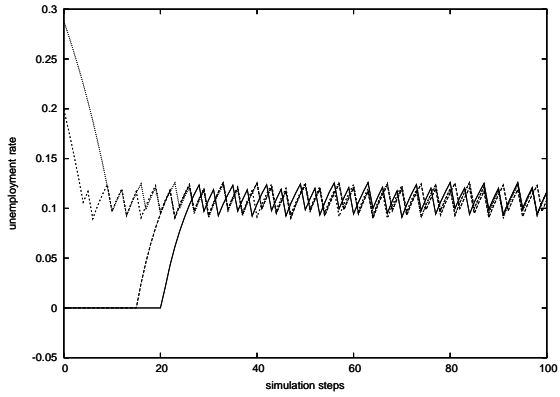}
\caption{Unemployment rate $(1-N_m/N_u)$ as function of simulation steps. From top to bottom the initial urban shares are $0.9,~0.8,~0.3,~0.2$.}
\label{unempt}
\end{figure}

\begin{figure}[hbt]
\includegraphics[width=7.5cm]{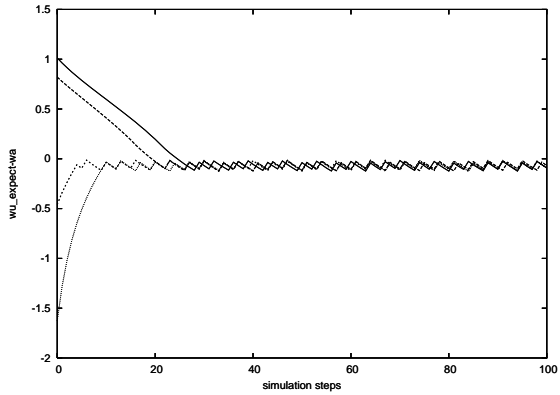}
\caption{Rural-urban expected wage differential $(w_u^e-w_a)$ as function of simulation steps. From top to bottom the initial urban shares are $0.9,~0.8,~0.3,~0.2$.}
\label{wu-wa}
\end{figure}

Figures \ref{nut}, \ref{unempt} and \ref{wu-wa} show the basic characteristics of the transitional dynamics and long run equilibrium generated by simulations. When the economic system has a low initial urban share, $n_u=0.2$ or $n_u=0.3$, there is a net migration toward urban sector. This migration takes the urban sector from a full employment situation to an unemployment one. The positive differential of expected wages that pulls workers to the urban sector diminishes. However, if the economic system initiates with a high urban share, $n_u=0.8$, or $n_u=0.9$ there is net flow of migration toward rural sector in such a way that the unemployment rate of the urban sector decreases. In this case, the differential of expected wages is negative.

In an economy mainly rural ($n_u<0.5$), the transitional dynamics characterized by a continuous growth of population of the urban sector with a differential of expected wages relatively high is followed by the stabilization of rural-urban differential of expected wages. In other words, the generalized Harris-Todaro condition, eq. (\ref{wuwadelta}), arises as a long run equilibrium result of the agent-based migratory dynamics.

Figure \ref{nut} also shows that even after the urban share has reached an stable average value, there are small fluctuations around this average. Therefore, differently from the original Harris-Todaro model, our computational model shows in the long run equilibrium the reverse migration. This phenomenon has been observed in several developing countries as remarked in Ref. \cite{day}.

\begin{figure}
\includegraphics[width=8.5cm]{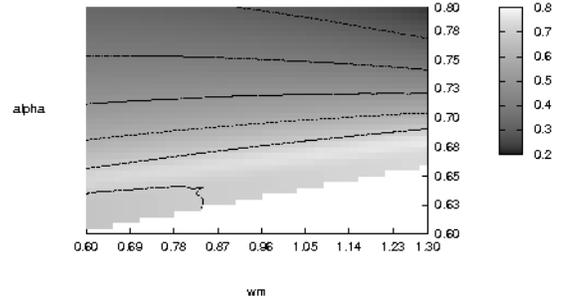}
\caption{Equilibrium urban share $n_u$ as function of the technological parameter $\alpha$ and the minimum wage $w_m$. White area is not a valid combination of parameters.}
\label{wmalphanu}
\end{figure}

\begin{figure}
\includegraphics[width=8.5cm]{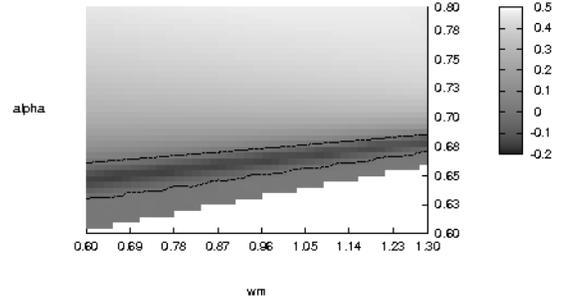}
\caption{Equilibrium differential of expected wages as function of the technological parameter $\alpha$ and the minimum wage $w_m$. White area is not a valid combination of parameters.}
\label{wmalphawuwa}
\end{figure}

\begin{figure}
\includegraphics[width=8.5cm]{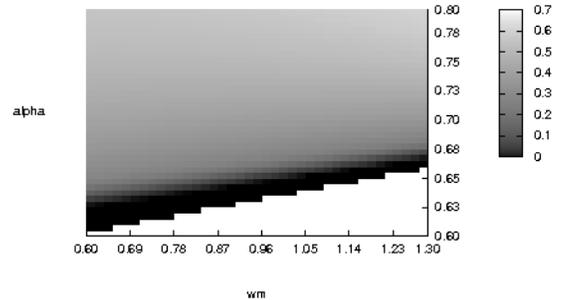}
\caption{Equilibrium urban unemployment rate $(1-N_m/N_u)$ as function of the technological parameter $\alpha$ and the minimum wage $w_m$. White area is not a valid combination of parameters.}
\label{wmalphaunemp}
\end{figure}

In Figures \ref{wmalphanu}, \ref{wmalphawuwa} and \ref{wmalphaunemp} one can see that for a given value of $\alpha$, the variation of $w_m$ practically does not change the equilibrium values of the urban share, the differential of expected wages and the unemployment rate. However, for a given $w_m$, higher values of $\alpha$ make the urban sector less attractive due the reduction of the employment level. This causes a lower equilibrium urban share, a higher unemployment rate and a gap in the convergence of the expected wages.

\begin{figure}
\includegraphics[width=8.5cm]{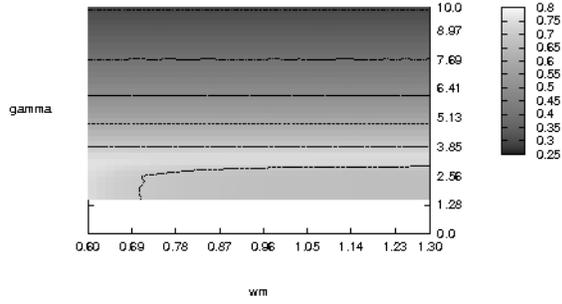}
\caption{Equilibrium urban share $n_u$ as function of the parameter $\gamma$ and the minimum wage $w_m$. White area is not a valid combination of parameters.}
\label{wmgammanu}
\end{figure}

\begin{figure}
\includegraphics[width=8.5cm]{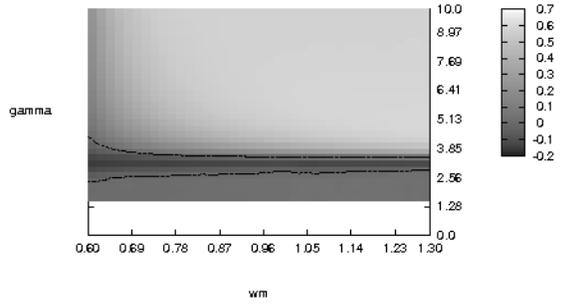}
\caption{Equilibrium differential of expected wages $(w_u^e-w_a)$ as function of the parameter $\gamma$ and the minimum wage $w_m$. White area is not a valid combination of parameters.}
\label{wmgammawuwa}
\end{figure}

\begin{figure}
\includegraphics[width=8.5cm]{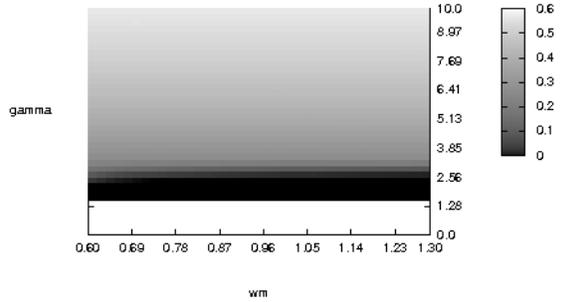}
\caption{Equilibrium urban unemployment rate $(1-N_m/N_u)$ as function of the parameter $\gamma$ and the minimum wage $w_m$. White area is not a valid combination of parameters.}
\label{wmgammaunemp}
\end{figure}

In Figures \ref{wmgammanu}, \ref{wmgammawuwa} and \ref{wmgammaunemp} can be seen that for a fixed value of $\gamma$, the equilibrium values of the urban share, the differential of expected wages and unemployment rate do not have a strong dependence with $w_m$. However, variations in $\gamma$ for a fixed $w_m$, dramatically change the equilibrium values of the variable mentioned before. Higher values of $\gamma$ generate a lower urban concentration, a higher gap in the expected wages and a higher unemployment rate in the equilibrium.

\begin{figure}
\includegraphics[width=8.5cm]{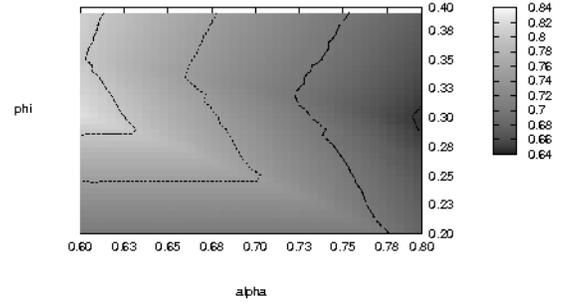}
\caption{Equilibrium urban share $n_u$ as function of the technological parameters $\alpha$ and $\phi$. White area is not a valid combination of parameters.}
\label{alphaphinu}
\end{figure}

Finally, in Figure \ref{alphaphinu} is shown that the convergence of migratory dynamics for a urban share, compatible with historical data, is robust in relation to the variation of the key technological parameters, $\alpha$ and $\phi$.

\begin{figure}
\includegraphics[width=8.5cm]{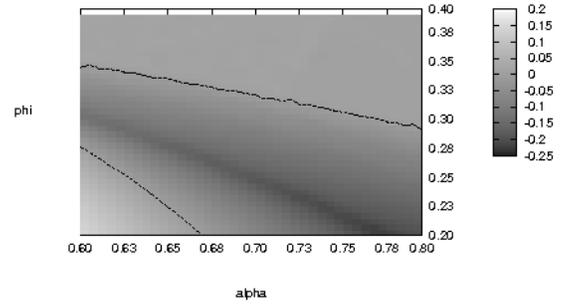}
\caption{Equilibrium differential of expected wages $(w_u^e-w_a)$ as function as function of the technological parameters $\alpha$ and $\phi$. White area is not a valid combination of parameters.}
\label{alphaphiwuwa}
\end{figure}

\begin{figure}
\includegraphics[width=8.5cm]{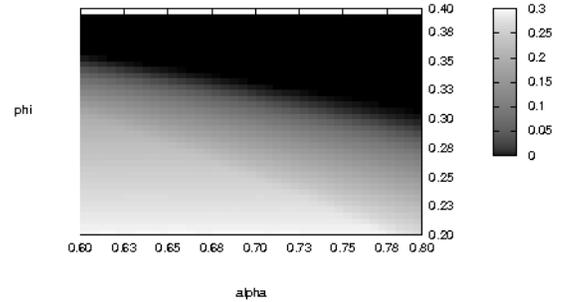}
\caption{Equilibrium urban unemployment rate $(1-N_m/N_u)$ as function of the technological parameters $\alpha$ and $\phi$. White area is not a valid combination of parameters.}
\label{alphaphiunemp}
\end{figure}

\section{Conclusion}
\label{conclusion}

In this paper we developed and agent-based computational model which formalizes the rural-urban allocation of workers as a process of social learning by imitation. We analyze a two-sectorial economy composed by adaptative agents, i.e., individuals that grope over time for best sectorial location in terms of earnings. This search is a process of imitation of successful neighbor agents.

The dispersed and non-coordinated individual migration decisions, made based on local information, generate aggregate regularities. Firstly, the {\it crucial assumption} of Harris and Todaro, the principle that rural-urban migration will occur while the urban expected wage exceed the rural wage, comes out as spontaneous upshot of interaction among adaptative agents. 

Secondly, the migratory dynamics generated by agents that seek to adaptate to the economic environment that they co-create leads the economy toward a long run equilibrium characterized by urban concentration with urban unemployment. When this long run equilibrium is reached, the generalized Harris-Todaro condition is satisfied, i.e., there is a stabilization of the rural-urban expected wage differential.

Thirdly, the impact of the minimum wage and elasticity of terms of trade in a long run equilibrium obtained by simulations are in agreement with the predictions of the original Harris-Todaro model with Cobb-Douglas technology.

Finally, the simulations showed an aggregated pattern not found in the original Harris-Todaro model. There is the possibility of small fluctuations of the urban share around an average value. This phenomenon is known as reverse migration.

\section*{Acknowledgments}

 Aquino L. Esp\'{\i}ndola thanks CAPES for the financial support. Jaylson J. Silveira acknowledges research grants from CNPq. T. J. P. Penna thanks CNPq for the fellowship.

\end{document}